\documentclass[12pt,dvips]{article}
\usepackage{cite}
\usepackage{bm}
\usepackage{graphicx}
\usepackage{color}
\begin{document}

\baselineskip 0.7cm

\begin{titlepage}

\begin{flushright}
IPMU17-0019\\
\end{flushright}

\vskip 1.35cm
\begin{center}
{\large \bf
Dirac CP phase in the neutrino mixing matrix and \\the Froggatt-Nielsen mechanism with ${\rm \bf det} \bf [M_\nu]=0$
}
\vskip 1.2cm
~Yuya~Kaneta$^{a}$,\ 
Morimitsu Tanimoto$^b$
~and~
Tsutomu T. Yanagida$^c$
\vskip 0.4cm
$^a${\it \normalsize
Graduate~School~of~Science~and~Technology,~Niigata University, \\
Niigata~950-2181,~Japan \\} 
$^b${\it \normalsize 
Department of Physics, Niigata University, Niigata 950-2181, Japan \\}
$^c${\it \normalsize 
Kavli Institute for the Physics and Mathematics of the Universe, \\ University of Tokyo, Kashiwa 277-8583, Japan \\}

\vskip 1.5cm

\abstract{
We  discuss  the Dirac CP violating phase $\delta_{CP}$
in the  Froggatt-Nielsen model 
for a neutrino mass matrix $M_\nu$  imposing  a condition ${\rm  det} [M_\nu]=0$. 
This additional condition restricts  the CP violating phase  $\delta_{CP}$ drastically.
We find  that the phase $\delta_{CP}$ is predicted in the region of  $\pm (0.4- 2.9)$ radian,
which is consistent with the recent T2K and NO$\nu$A data.
There is  a remarkable  correlation  between  $\delta_{CP}$ and $\sin^2\theta_{23}$.
The  phase $\delta_{CP}$ converges to  $\sim \pm \pi/2$ if  $\sin^2\theta_{23}$  
is  larger than $0.5$.
Thus,  the accurate measurement of  $\sin^2\theta_{23}$ is important for a test of our model.
The  effective mass $m_{ee}$  for  the  neutrinoless  double beta decay is predicted 
in the rage  $3.3-4.0$ meV.
}
\end{center}
\end{titlepage}

\setcounter{page}{2}

\section{Introduction}
The  Froggatt-Nielsen (FN) mechanism  \cite{Froggatt:1978nt} is very attractive since it naturally explains the observed masses and mixing angles for
quarks and leptons. It is well known that  the magnitudes of observed mixing angles for quarks are given by  powers of  Wolfenstein parameter 
$\lambda  \simeq  0.2$ \cite{Wolfenstein:1983yz}. 
This is nothing but the feature predicted by the FN mechanism. 
The  lepton  flavor  mixing matrix, so called MNS matrix
\cite{Maki:1962mu,Pontecorvo:1967fh},   exhibits two large mixing angles, 
and one rather small mixing angle of the  order of Cabibbo angle.
Surprisingly, this lepton mixing matrix is also explained by the FN mechanism 
\cite{Buchmuller:1998zf,Vissani:1998xg,Vissani:2001im,Sato:2000kj,Elwood:1998kf,Ling:2002nj,Bando:2001bj}.

Among various proposals, Ling and Ramond \cite{Ling:2002nj} 
presented a clear phenomenological discussion of neutrino masses and mixing angles
in terms of Cabibbo angle $\lambda_C \simeq 0.225$ \cite{Olive:2016xmw}.
Their texture is still consistent with  the  recent precise data on the three  neutrino mixing angles
and two neutrino mass squared differences.
However, this texture cannot predict the CP phase as we discuss later in this paper.

The neutrino oscillation experiments are now on a new step to confirm the CP violation in the lepton sector.
Actually, the T2K and NO$\nu$A experiments indicate a finite CP phase 
\cite{Abe:2014ugx,Abe:2017uxa,Adamson:2016tbq,Bian:2016iga}. Therefore, 
it is very interesting to extend the FN model to predict the Dirac CP violating phase.

In this paper, we discuss the Dirac  CP violating phase in 
the FN model for  the neutrino mass matrix $M_\nu$ imposing  an additional  condition ${\rm  det} [M_\nu]=0$ \cite{Branco:2002ie}. 
This flavor-basis independent condition of ${\rm  det} [M_\nu]=0$ is obtained easily 
by assuming  two families of heavy right-handed neutrinos \cite{Frampton:2002qc}
in the framework of the seesaw mechanism  \cite{Minkowski,GRSY}.
It is also interesting that the Affleck-Dine scenario  \cite{Affleck:1984fy} 
for leptogenesis \cite{Murayama:1993em,Dine:1995kz} 
requires the mass of the lightest neutrino to be $m_1=10^{-10}$ eV  \cite{Asaka:2000nb,Fujii:2001sn},  which  practically leads to our condition ${\rm  det} [M_\nu]=0$.
We show that the phase $\delta_{CP}$ is predicted in a narrow region  using the presently
available data on the mass squared differences and the  mixing angles.

In section 2, we discuss a texture of the neutrino mass matrix  imposing ${\rm  det} [M_\nu]=0$ in the
FN model, where neutrinos are supposed to be Majorana particles.
In section 3, we show numerical results on  $\delta_{CP}$, $\sin^2\theta_{23}$, $\sin^2\theta_{13}$ and
$\sin^2\theta_{12}$. The effective mass $m_{ee}$ that appears in the neutrinoless double 
beta decay is also discussed.
The  summary and discussion are  given in section 4.

\section{FN texture for leptons}
Let us discuss lepton mass matrices in the framework of the FN model.
We assign the  FN charges of the FN  broken U$(1)$ \cite{Froggatt:1978nt}
to the three left-handed leptons $\ell_{Li}$ as
\begin{equation}
 {\ell_{L1}, \ell_{L2},\ell_{L3}}\ :\ {n+1, n,n} \ ,
\end{equation}
where $n$ is a positive integer.
Then, the mass matrix of the left-handed Majorana neutrinos is given in terms of
the FN parameter $\lambda$,  which is of the order of the Cabibbo angle $\lambda_C$, as follows: 
\begin{eqnarray}
  M_{\nu} \sim \left( \matrix{\lambda^2 &  \lambda & \lambda \cr   \lambda & 1 & 1 \cr
    \lambda & 1 & 1 \cr} \right )\  . 
  \label{Mnu}
\end{eqnarray}
This mass matrix leads to  the Normal Hierarchy (NH) of neutrino masses, and
gives us evidently one large mixing angle between the second and third families of neutrinos. 
Namely, the  FN charge of the  left-handed leptons is chosen by the observed large mixing.
The charged lepton mass matrix is given after fixing  FN charges of the right-handed charged leptons
to reproduce the observed mass hierarchy among the charged leptons.
Assigning   FN charges to the three right-handed charged leptons $e_{Ri}$ as
\begin{equation}
  {e_{R1}, e_{R2}, e_{R3}}\ :\ {4, 2, 0} \ ,
\end{equation}
the charged lepton mass matrix $M_{E}$ is given as 
\begin{eqnarray}
  M_{E} \sim \left( \matrix{\lambda^5 & \lambda^3 & \lambda \cr   \lambda^4 & \lambda^2 &1\cr
    \lambda^4 & \lambda^2 & 1 \cr   } \right )\  , 
\end{eqnarray}
which gives the  mass ratio in terms of $\lambda$ as follows:
\begin{equation}
  \frac{m_e}{m_\tau}\sim \lambda^5 \ , \qquad  \frac{m_\mu}{m_\tau}\sim \lambda^2 \ .
  \label{massratio}
\end{equation}
These mass ratios are consistent with observed ones for about  $\lambda\simeq 0.2$.

We move to  the diagonal basis of the charged lepton mass matrix
in order to reduce the number of free parameters.
Then, the rotation of the left-handed lepton doublets to diagonalize the charged lepton mass matrix
does not change the  powers of $\lambda$ in the entries of the  neutrino mass matrix  of  Eq.(\ref{Mnu}).
Therefore, we discuss the following neutrino mass matrix in the diagonal basis of the charged lepton mass matrix:
\begin{eqnarray}
  M_{\nu} =m_0 \left( \matrix{a\lambda^2 & b \lambda & c\lambda \cr  b \lambda & d &e \cr
    c\lambda & e & f \cr   } \right )\  , 
\end{eqnarray}
where $a-f$ are dimensionless complex parameters with  their magnitudes of the order $1$
\footnote{Due to the rotation of the left-handed lepton doublets,
the magnitude of the coefficients $a-f$ may be rather enlarged. We address this point in the section 4.} .

By using the freedom of  phase redefinition of the left-handed lepton fields,
we take the diagonal elements to be real.
Then, we have three CP phases in the mass matrix.
Now, we parameterize the mass matrix in order to analyze the neutrino mixing numerically  as
\begin{eqnarray}
  M_{\nu} =m_0 \left( \matrix{a\lambda^2 & b \lambda e^{i\phi_b} & c \lambda e^{i\phi_c}\cr 
    b e^{i\phi_b} \lambda & d &e e^{i\phi_e} \cr
    c\lambda e^{i\phi_c} & e e^{i\phi_e} & 1 \cr   } \right )\  , 
  \label{texture}
\end{eqnarray}
where $a-e$ are redefined as real parameters of the order $1$.

Let us determine the magnitude of $\lambda$ from the  observed charged lepton mass ratios
in Eq.(\ref{massratio}).
We use the $m_e/m_\tau$ ratio to fix $\lambda$, since it has the strongest $\lambda$
dependence among charged lepton mass ratios as seen in Eq.(\ref{massratio}). Then, 
we obtain $\lambda\simeq 0.20$ from the $m_e/m_\tau$ ratio. 
Taking  into account  the order one coefficients in  those mass ratios in  Eq.(\ref{massratio}),
$\lambda=0.20$ can explain  all lepton mass ratios consistently.
We take   $\lambda=0.18\sim 0.22$ considering the ambiguity of $10\%$ for $\lambda$
in our numerical computation 
\footnote{The ambiguity of the coefficients $a-e$ due to the rotation of the left-handed lepton doublets is partially absorbed by taking account of the  ambiguity of $10\%$ for $\lambda$.}.
We have now nine  parameters $m_0$, $a-e$, $\phi_b$, $\phi_c$ and $\phi_e$, where
the $m_0$ has dimension of a mass, but others are dimensionless parameters.

Let us impose a flavor-basis independent condition that the
determinant of the neutrino mass matrix vanishes, that is ${\rm  det} [M_\nu]=0$. This condition gives two constraints on the parameters, and then the neutrino mass matrix
has now just seven free parameters  which can be fully determined by future feasible experiments \cite{Branco:2002ie}. We see below that 
thanks to this condition, we can predict the CP violating phase  $\delta_{CP}$,
which is defined in the Particle Data Group \cite{Olive:2016xmw}.

\section{Numerical Analysis}
Let us present our numerical analysis of the neutrino mass matrix in Eq.(\ref{texture}).
The  free parameters  $a-e$ are of the order one. 
We scan them  in the region of $0.7\sim 1.3$ by generating random numbers in the liner scale.
Our choice of  the  parameter region of $0.7\sim 1.3$ is justified later by the predicted mixing of
$\sin^2\theta_{23}$. 
The parameter  $\lambda$ is essentially given by the FN model. As discussed in the previous section,
 the charged lepton mass hierarchy indicates $\lambda\simeq 0.2$.  In our numerical analysis,
 we also scan it at random with  the liner scale in the region $\lambda=0.18\sim 0.22$.
Furthermore,
the extension of the scanning region, for example, $0.5\sim 2$ is not favored
because the hierarchies between   $a\lambda^2$ and $b\lambda(c\lambda)$,
and between $b\lambda(c\lambda)$ and $d$,  are no longer distinguishable,
and then the FN scheme with $\lambda\simeq 0.2$ becomes meaningless.

The CP violating phases $\phi_b$, $\phi_c$ and $\phi_e$
are also  scanned in the full region of $-\pi\sim \pi$ by generating random numbers in the liner scale.

Now we explain how to obtain our predictions in our figures.
By scanning the parameters of  $a-e$ and three phases with $\lambda=0.18\sim 0.22$, 
we generate a neutrino mass matrix.  
The parameter $m_0$ is determined to reproduce
the observed values of $\Delta m_{23}^2$ and $\Delta m_{12}^2$ at  $2\sigma$ interval in Table 1.
In practice, $m_0$ is also scanned randomly in the linear scale up to 
   the upper bound of the total neutrino mass $0.2$ eV,
    which is given by the cosmology observation  \cite{Olive:2016xmw}.
   Actually,  the obtained $m_0$ is in the region of  $ (0.025-0.035)$ eV.
 It is noticed,  in the case of ${\rm  det} [M_\nu]=0$, $m_0$ is easily determined by the experimental data 
 of $\Delta m_{12}^2$ and $\Delta m_{23}^2$  because of  $m_1=0$.

  Then, we obtain the calculated three mixing angles.
  If these predicted  mixing angles are OK for the experimental data in Table 1, 
we keep the point. Otherwise, we disregard the point.
  We continue this procedure to obtain $10^4$ points, which satisfy the experimental data.

\begin{table}[hbtp]
\begin{center}
\begin{tabular}{|c|c|c|}
\hline 
\  observable \ &  best fit and $1\sigma$ & $2\sigma$ interval \\
\hline 
 $\Delta m_{23}^2$& \ \   $2.524^{+0.039}_{-0.040} \times 10^{-3}{\rm eV}^2 $ \ \
                             &\ \ $(2.44\sim 2.56) \times 10^{-3}{\rm eV}^2$ \ \  \\
\hline 
 $\Delta m_{12}^2$&  $7.50^{+0.19}_{-0.17} \times 10^{-5}{\rm eV}^2$ 
                            & $(7.16\sim 7.88)  \times 10^{-5}{\rm eV}^2$ \\
\hline
 $\sin^2\theta_{23}$&  $0.441^{+0.027}_{-0.021}$ & $0.39\sim 0.63$ \\
\hline 
$\sin^2\theta_{12}$& $0.306 \pm 0.012$ & $0.28\sim 0.33$ \\
\hline 
 $\sin^2\theta_{13}$&  $0.02166\pm 0.00075$ & $0.020\sim 0.023$ \\
\hline 
\end{tabular}
\caption{Results of the global analysis of the neutrino oscillation experimental data
  for NH of neutrino masses \cite{Esteban:2016qun}, where 
  observables are defined in Particle Data Group \cite{Olive:2016xmw}.}
\label{tab}
\end{center}
\end{table}

\subsection{Prediction of mixing angles}
\begin{figure}[h]
\begin{minipage}[]{0.45\linewidth}
\includegraphics[width=8cm]{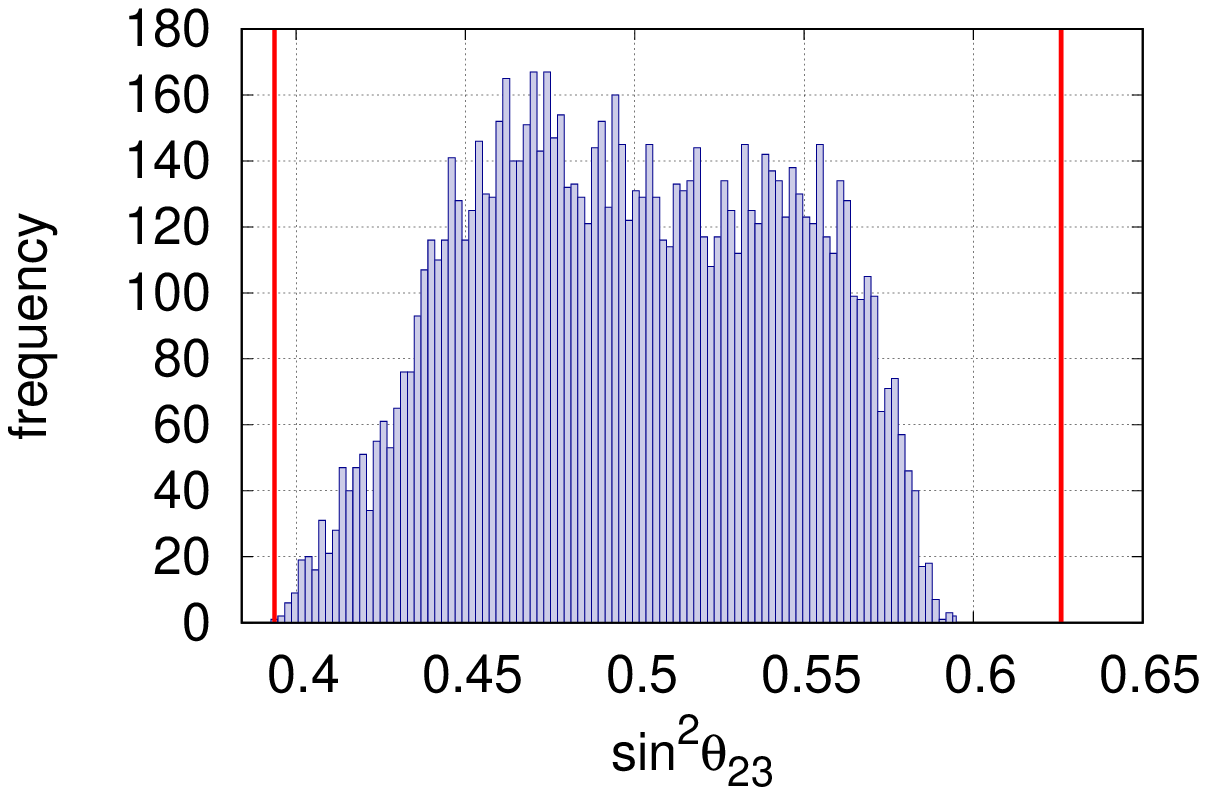}
\caption{The frequency distribution of the predicted $\sin^2\theta_{23}$
  under the condition of  ${\rm  det} [M_\nu]=0$, where
  only the  experimental data of $\Delta m^2_{23}$ and  $\Delta m^2_{12}$ are input.
  The vertical red lines denote the observed  $\sin^2\theta_{23}$  interval  at $2\sigma$.}
\end{minipage}
\hspace{5mm}
\begin{minipage}[]{0.45\linewidth}
\includegraphics[width=8cm]{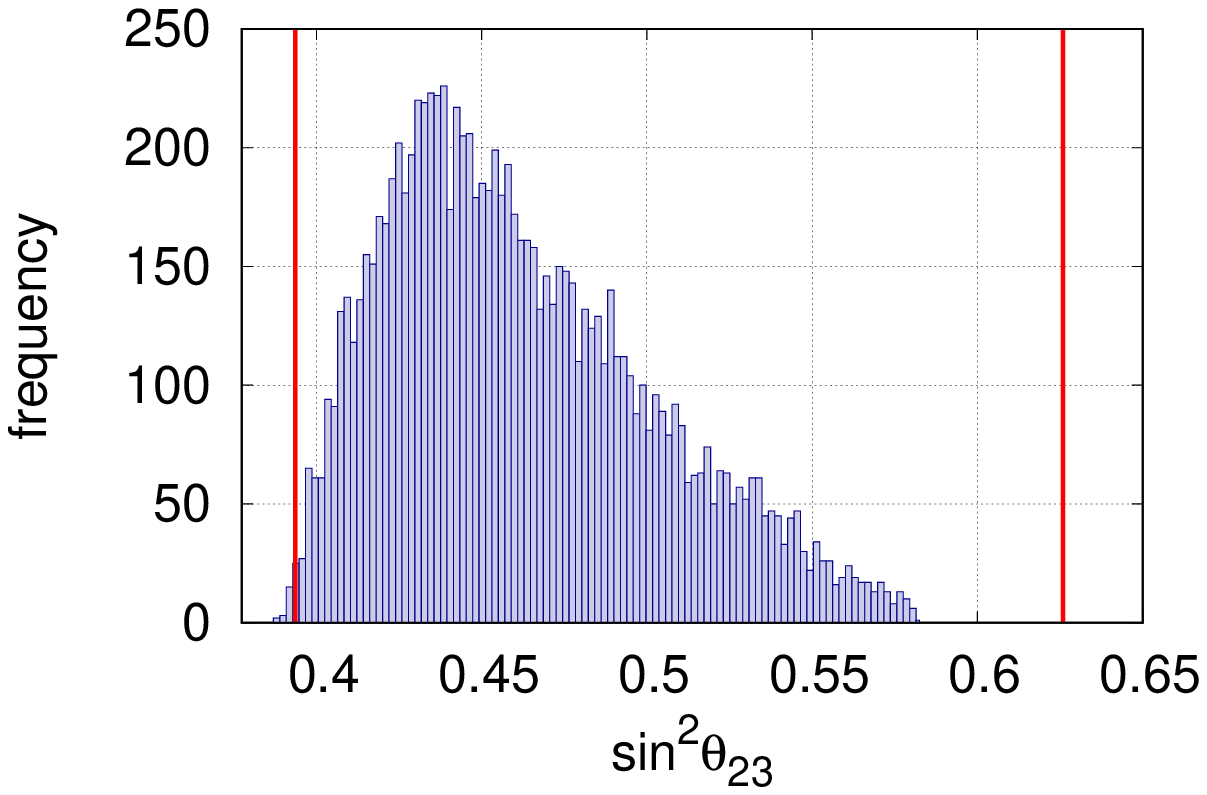}
\caption{The frequency distribution of the predicted $\sin^2\theta_{23}$
  with adding the input data of $\sin^2\theta_{12}$ with  $2\sigma$ error-bar, 
  where the condition ${\rm  det} [M_\nu]=0$ is imposed.
  The vertical red lines denote the observed  $\sin^2\theta_{23}$  interval  at $2\sigma$.}
\end{minipage}
\end{figure}
First, we discuss the  mixing angle $\theta_{23}$  by imposing ${\rm  det} [M_\nu]=0$.
This mass matrix leads to a large mixing angle $\theta_{23}$ naturally
since  all elements  of the submatrix for the second and third families are of  the order one.
We can predict the magnitude of $\sin^2\theta_{23}$ by using only the experimental data $\Delta m^2_{23}$ and  $\Delta m^2_{12}$ with   $2\sigma$ error-bar in Table 1.
We show the  frequency distribution of the predicted $\sin^2\theta_{23}$  in Fig. 1, 
where  $\lambda=0.18\sim 0.22$ and $a-e=0.7\sim 1.3$  are taken.
It is remarkable that predicted  $\sin^2\theta_{23}$ lies inside  the experimental allowed region
of $2\sigma$. 
The prediction almost distributes around $0.5$  symmetrically.
The predicted region of  $\sin^2\theta_{23}$
depends on our choice of   $a-e=0.7\sim 1.3$.
That is to say, our choice of  $a-e=0.7\sim 1.3$  nicely predicts $\sin^2\theta_{23}$
for the  fixed   $\lambda=0.18\sim 0.22$.
For example, an extension of the scanning region such as $a-e=0.5\sim 1.5$
leads to $\sin^2\theta_{23}$ which lies over the experimental allowed region.
This is a reason why we take $a-e=0.7\sim 1.3$ in this paper.

Let us  use the constraint from the data $\sin^2\theta_{12}$ 
with  $2\sigma$ error-bar in Table 1 in addition to the data of  $\Delta m^2_{23}$ and  $\Delta m^2_{12}$.
The  predicted  $\sin^2\theta_{23}$ is shown  in Fig.2.
The frequency distribution  of $\sin^2\theta_{23}$ is remarkably  changed.
It is asymmetric around $0.5$ as seen in Fig.2.  The region $\sin^2\theta_{23}<0.5$ is favored.
It may be interesting to comment that this prediction is not changed 
even if  the data of $\sin^2\theta_{13}$  is added. 
Thus, the input of $\sin^2\theta_{12}$  pushes  $\sin^2\theta_{23}$ toward a region smaller 
than $0.5$.
It is interesting that the peak of the frequency distribution is  around $0.44$, which is the 
best fit value of the experimental data as seen in Table 1.

We add a comment that the distribution plot of Fig.2 
 covers  all region of the experimental interval of   $\Delta m^2_{23}$ and  $\Delta m^2_{12}$ in Table 1.
 It also covers  all region of  the experimental interval of  $\sin^2\theta_{12}$  as seen later in Fig. 8.

\begin{figure}[h]
\begin{minipage}[]{0.45\linewidth}
\includegraphics[width=8cm]{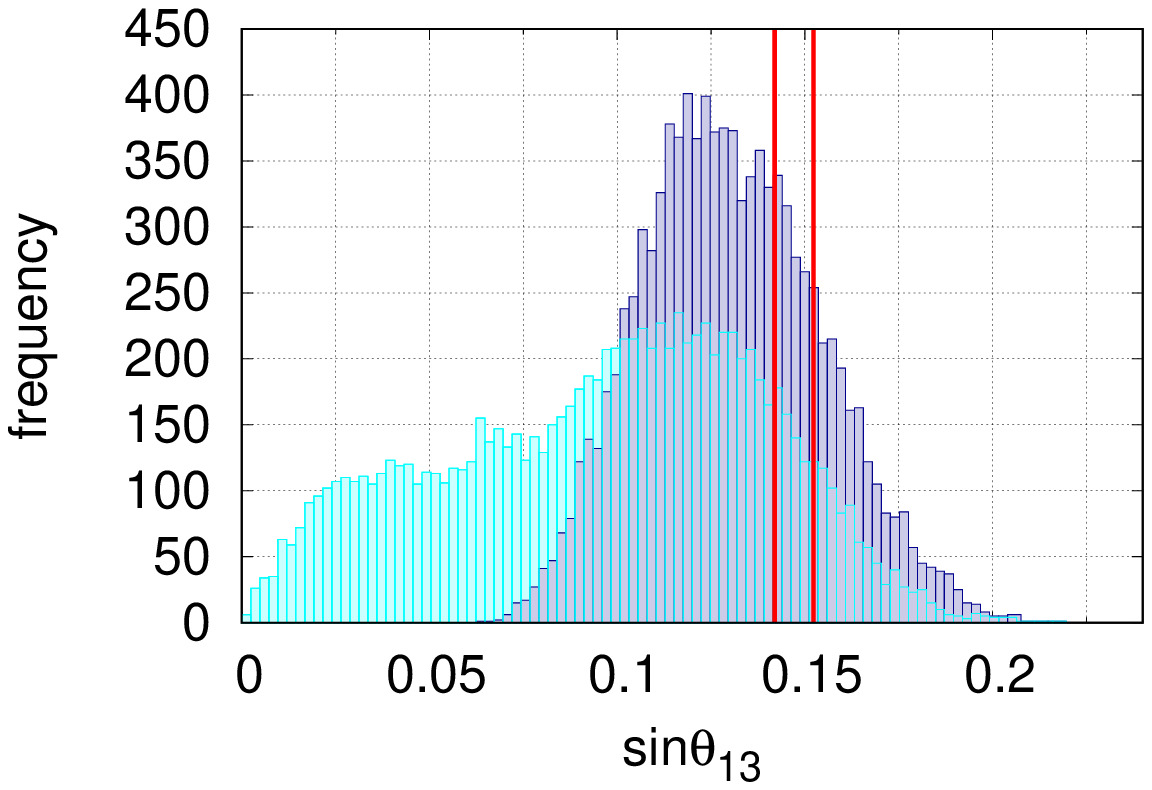}
\caption{The frequency distribution of the predicted $\sin\theta_{13}$,
where the blue (cyan) corresponds to  the case with (without) ${\rm  det} [M_\nu]=0$.
Only the  experimental data of $\Delta m^2_{23}$ and  $\Delta m^2_{12}$ are used as inputs.
The vertical red lines denote the observed  $\sin\theta_{13}$  interval  at $2\sigma$.}
\end{minipage}
\hspace{5mm}
\begin{minipage}[]{0.45\linewidth}
\includegraphics[width=8cm]{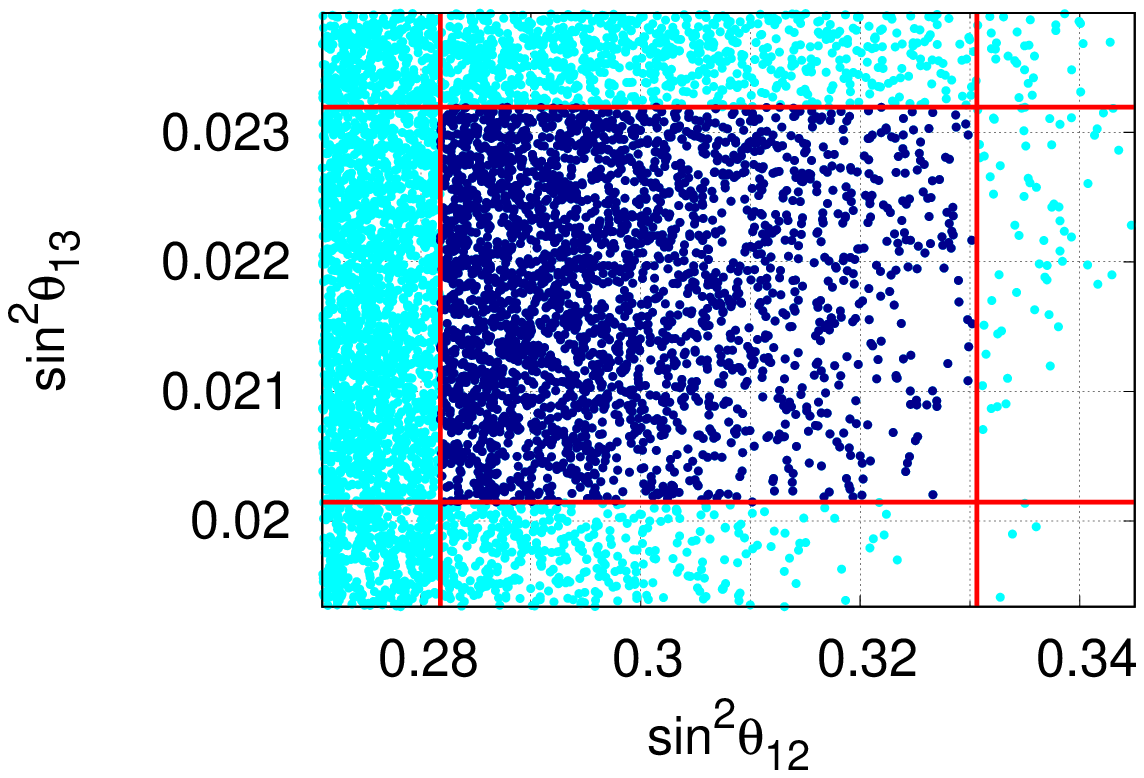}
\caption{The predicted region  on the plane of $\sin^2\theta_{12}$ and $\sin^2\theta_{13}$
  under the condition of  ${\rm  det} [M_\nu]=0$, where
  only the  experimental data of $\Delta m^2_{23}$ and  $\Delta m^2_{12}$ are used as inputs.
  The scattered plot is  shown in the experimental allowed region with $3\sigma$.
  The red lines denote the observed interval at $2\sigma$.}
\end{minipage}
\end{figure}

The other mixing angles  $\theta_{12}$ and $\theta_{13}$ are also predictable.
We show the  frequency distribution of the predicted $\sin \theta_{13}$ 
with/without imposing ${\rm  det} [M_\nu]=0$ in Fig. 3, where
only the  experimental data of $\Delta m^2_{23}$ and  $\Delta m^2_{12}$ are used as inputs.
The tiny $\sin\theta_{13}$ is still allowed in spite of the $(1,3)$ matrix element being of
the order $\lambda$ in Eq.(\ref{texture}) unless the condition of ${\rm  det} [M_\nu]=0$ is  imposed.
It is remarkable that the condition of ${\rm  det} [M_\nu]=0$  excludes 
the smaller region than $0.07$ for $\sin\theta_{13}$ as seen in Fig.3.
Thus, the condition  of ${\rm  det} [M_\nu]=0$ leads to $\sin \theta_{13}\simeq 0.1$ naturally.

On the other hand,  the predicted region of $\theta_{12}$ is rather broad.
It is understandable that   the $(1,2)$ entry of  the neutrino mass matrix  in Eq.(\ref{texture})
could be  drastically reduced after the large  rotation of the second and third family axes
since both $(1,2)$ and $(1,3)$ entries  are of  the order $\lambda$.  In particular, a large cancellation
in the $(1,2)$ entry is required to satisfy the condition  ${\rm  det} [M_\nu]=0$, since the $(3,3)$ entry
is much larger than the $(2,2)$ entry after the large rotation. In fact,
the predicted region  for  $\sin^2\theta_{12}$ contains  the region around $0$.
We present   the predicted region  on the plane of $\sin^2\theta_{12}$ and $\sin^2\theta_{13}$
with the condition of ${\rm  det} [M_\nu]=0$ in Fig. 4,
where the scattered plot is  shown in the experimental allowed region with $3\sigma$.
It is concluded that the predicted $\theta_{12}$ and $\theta_{13}$ are 
completely consistent with the experimental data.
Now, we try to predict  the CP violating phase $\delta_{CP}$ in the next subsection.

\begin{figure}[t]
\begin{minipage}[]{0.45\linewidth}
\includegraphics[width=8cm]{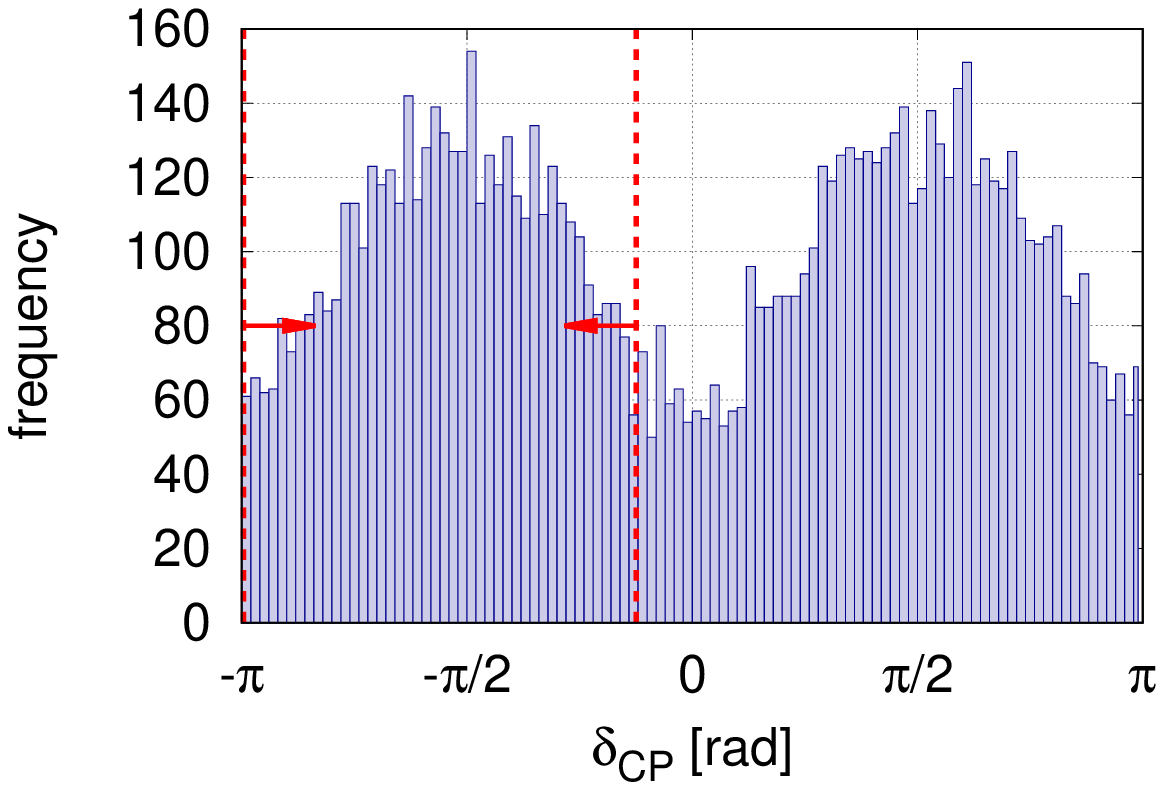}
\caption{The frequency distribution of the predicted $\delta_{CP}$ without  ${\rm  det} [M_\nu]=0$.
  The vertical dashed lines denote the observed $\delta_{CP}$ interval  at $90\%$ C.L.
  in the  T2K experiment  \cite{Abe:2017uxa}.}
\end{minipage}
\hspace{5mm}
\begin{minipage}[]{0.45\linewidth}
\includegraphics[width=8cm]{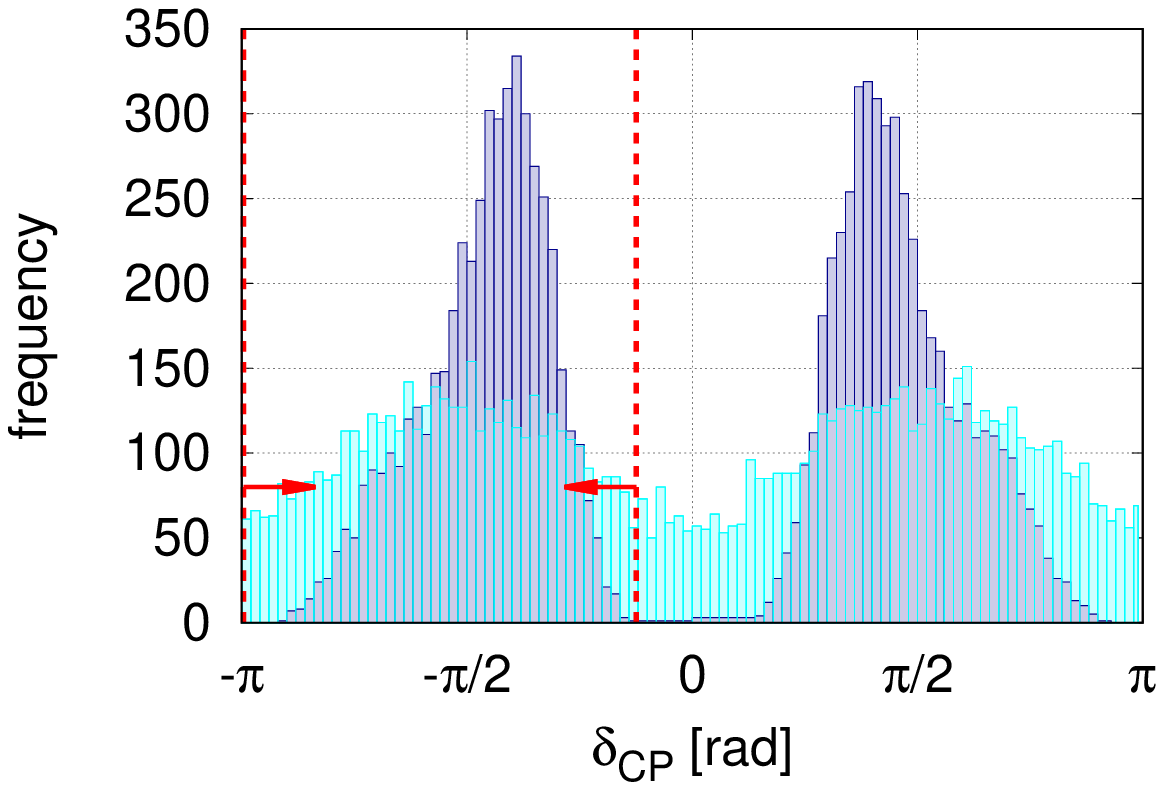}
\caption{The frequency distribution of the predicted $\delta_{CP}$, where 
  the blue (cyan) corresponds to  the case with (without) ${\rm  det} [M_\nu]=0$.
  The vertical  dashed lines denote the observed $\delta_{CP}$ interval  at $90\%$ C.L.
  in the  T2K experiment  \cite{Abe:2017uxa}.}
\end{minipage}
\end{figure}

\subsection{Prediction of $\delta_{CP}$}
In order to predict the CP violating phase $\delta_{CP}$ precisely, we  also use the data of  all mixing angles, $\theta_{23}$, $\theta_{12}$ and  $\theta_{13}$,
in addition to $\Delta m^2_{23}$ and  $\Delta m^2_{12}$.
At first, we show  the calculated frequency distribution of  $\delta_{CP}$  
without imposing ${\rm  det} [M_\nu]=0$ in Fig.5. 
The vertical dashed lines denote the observed $\delta_{CP}$ interval  at $90\%$ C.L.
in the recent T2K experiment  \cite{Abe:2017uxa}.
We see that the predicted $\delta_{CP}$ lies in the all region $-\pi\sim \pi$.

However, when ${\rm  det} [M_\nu]=0$ is imposed on  the neutrino mass matrix in Eq.(\ref{texture}),
$\delta_{CP}$ is  predicted   around $\pm \frac{\pi}{2}$ as seen in Fig.6,
where blue (cyan) corresponds to  the case with (without)  ${\rm  det} [M_\nu]=0$.
The CP conserved case   $\delta_{CP}=0,\pm\pi$ is excluded.
The allowed region of $\delta_{CP}$ is $\pm (0.4\sim 2.9)$ radian,
which is consistent with the observed $\delta_{CP}$ interval $-(0.39\sim 3.13)$ radian at $90\%$ C.L. by using 
the  Feldman-Cousins method for NH in the recent T2K experiment  \cite{Abe:2017uxa}.
Thus, the condition of ${\rm  det} [M_\nu]=0$ is essential for the  prediction of  $\delta_{CP}$.

We also discuss the correlations among mixing angle $\theta_{23}$ and CP violating phase $\delta_{CP}$.
We show the plot   $\delta_{CP}$ versus  $\sin^2\theta_{23}$ in Fig.7, where   ${\rm  det} [M_\nu]=0$
is imposed.
As  $\sin^2\theta_{23}$ increases, the predicted range of $\delta_{CP}$ becomes narrow.
If  $\sin^2\theta_{23}$  is  larger than $0.5$,  $\delta_{CP}$ converges toward $\pm \pi/2$.
Actually, the allowed region of $\delta_{CP}$ is $\pm (0.7\sim 2.4)$ radian.
More  accurate measurements of  $\sin^2\theta_{23}$ will be important to test our model.

We show the allowed region in the plane of $\sin^2\theta_{12}$ and $\sin^2\theta_{23}$ in Fig.8,
where  ${\rm  det} [M_\nu]=0$ is imposed.
The region where both of $\sin^2\theta_{12}$ and $\sin^2\theta_{23}$ are large is excluded.

\begin{figure}[h]
\begin{minipage}[]{0.45\linewidth}
\includegraphics[width=8cm]{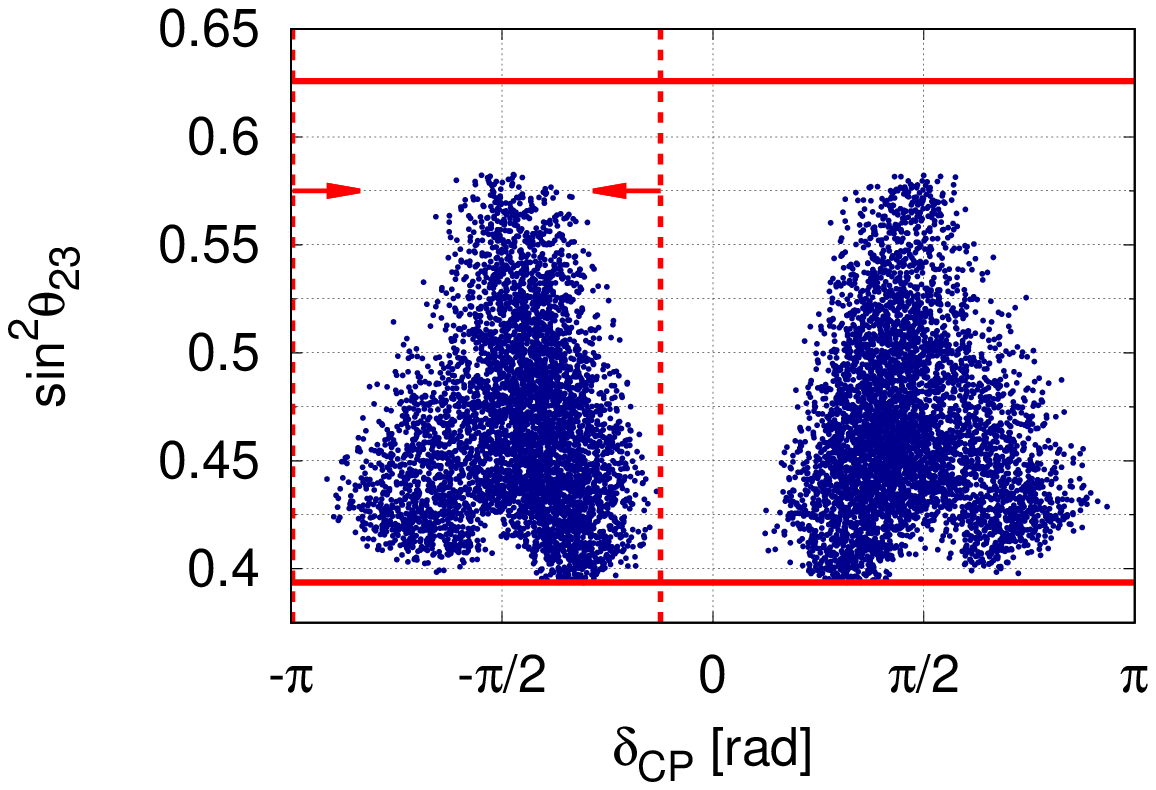}
\caption{The predicted $\delta_{CP}$ versus $\sin^2\theta_{23}$
  with imposing  ${\rm  det} [M_\nu]=0$.
  The horizontal red  lines denote the experimental bounds for $\sin^2\theta_{23}$ with  $2\sigma$.
  The vertical  dashed lines denote the observed $\delta_{CP}$ interval  at $90\%$ C.L.
  in the  T2K experiment  \cite{Abe:2017uxa}. }
\end{minipage}
\hspace{5mm}
\begin{minipage}[]{0.45\linewidth}
\includegraphics[width=8cm]{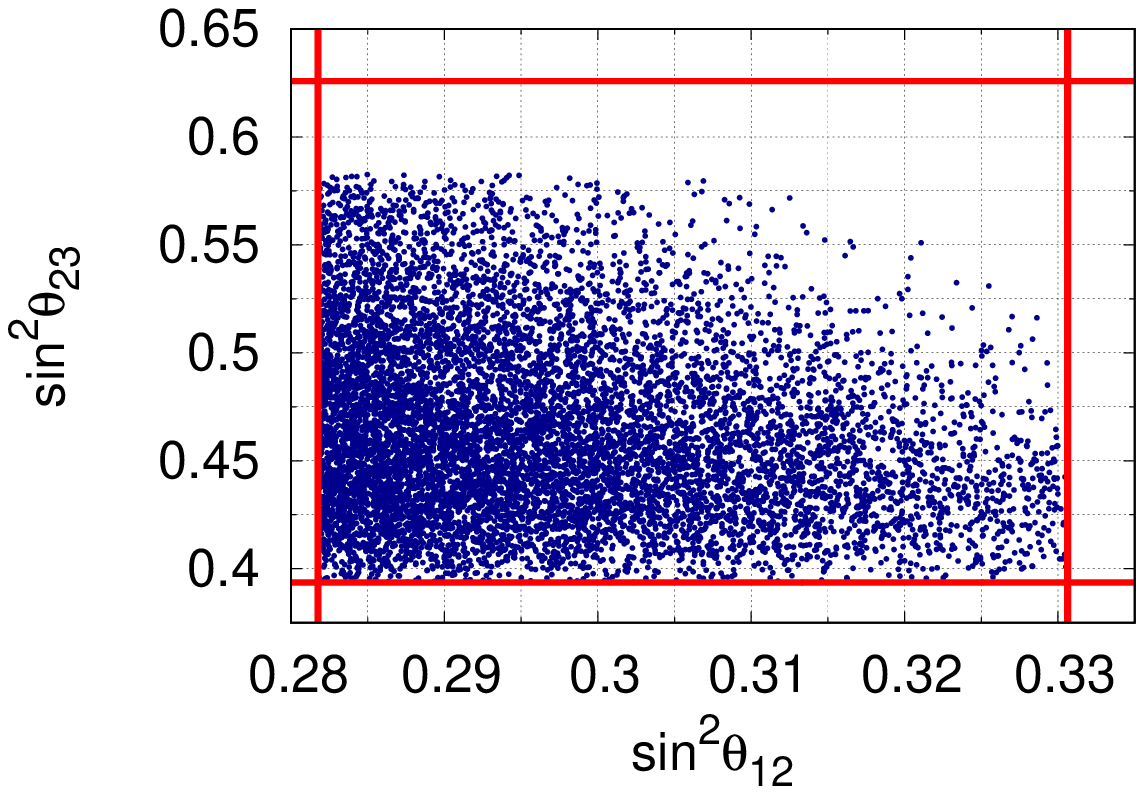}
\caption{The allowed region in the plane of $\sin^2\theta_{12}$ and $\sin^2\theta_{23}$
  with imposing  ${\rm  det} [M_\nu]=0$. The red  lines denote the experimental bounds 
  for  $\sin^2\theta_{12}$ and $\sin^2\theta_{23}$ with  $2\sigma$.}
\end{minipage}
\end{figure}

\subsection{Prediction of the  effective mass $m_{ee}$}
Finally, we  discuss the effective neutrino mass responsible for  the neutrinoless double 
beta decay 
\begin{eqnarray}
m_{ee}=\left |\sum _{i=2}^3 m_i U_{ei}^2 \right|  ~ ,
\end{eqnarray}
where $U_{ei}$ denotes the MNS mixing matrix element. 
We show  the frequency distribution of the predicted   $m_{ee}$, 
which lies in the range $m_{ee}=3.35-4.00$ meV,  in Fig.9,
where   ${\rm  det} [M_\nu]=0$ is imposed. 
\begin{figure}[h]
\begin{minipage}[]{0.45\linewidth}
\includegraphics[width=8cm]{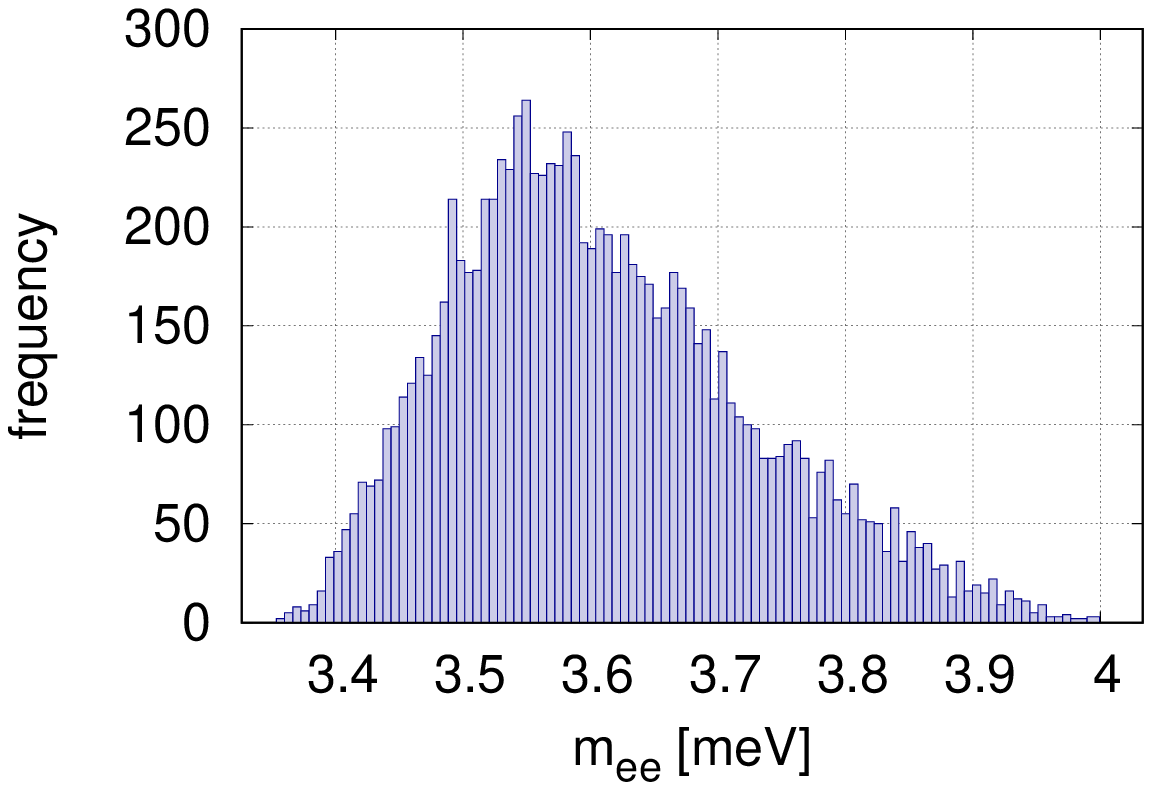}
\caption{The frequency distribution of the predicted $m_{ee}$  
  with imposing  ${\rm  det} [M_\nu]=0$.}
\end{minipage}
\hspace{5mm}
\begin{minipage}[]{0.45\linewidth}
\includegraphics[width=8cm]{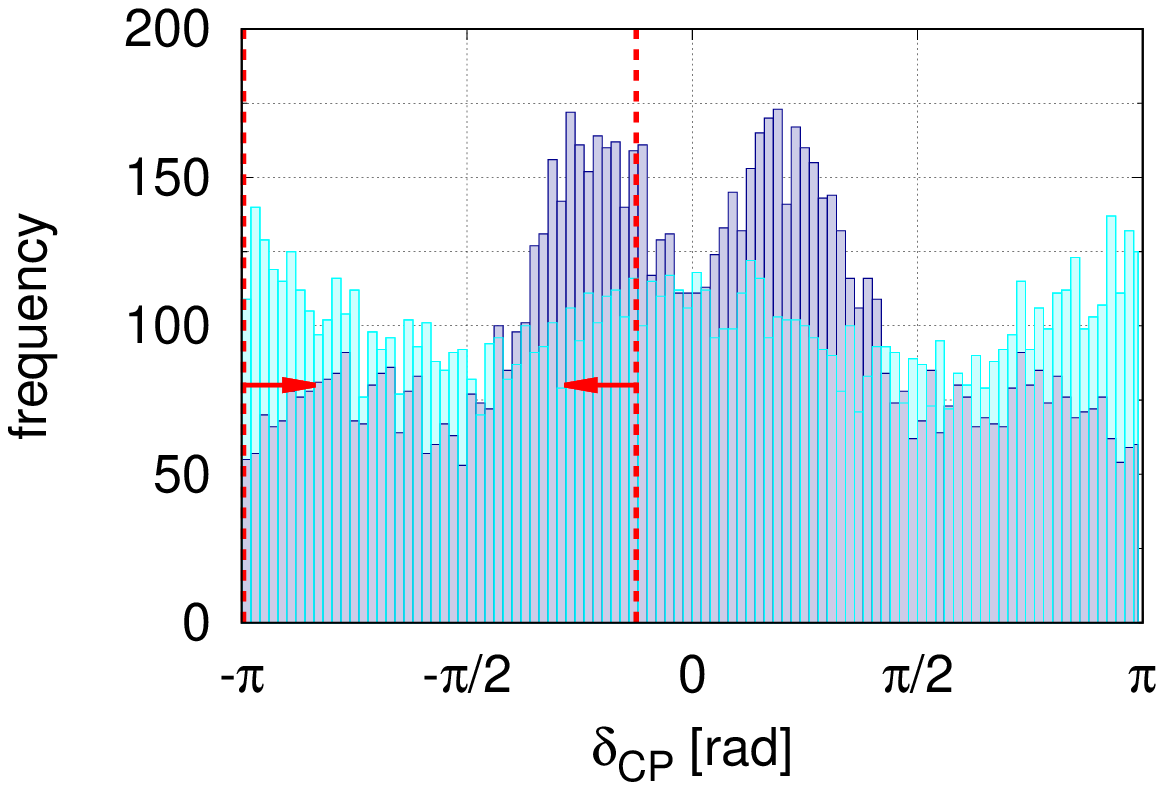}
\caption{The frequency distribution of the predicted $\delta_{CP}$ by scanning $a-e=0.5\sim 2$, where 
  the blue (cyan) corresponds to  the case with (without) ${\rm  det} [M_\nu]=0$.
  The vertical  dashed lines denote the observed $\delta_{CP}$ interval  at $90\%$ C.L.
  in the  T2K experiment.}
\end{minipage}
\end{figure}
\section{Summary and Discussion}
We have discussed the mixing angles  and the Dirac CP violating phase
in the framework of the FN model  with the  flavor-basis independent condition ${\rm  det} [M_\nu]=0$.
It is remarkable that   $\sin^2\theta_{23}$ is predicted inside of the experimental allowed region
of $2\sigma$, where we have used  only the data of   $\Delta m^2_{23}$ and  $\Delta m^2_{12}$.
Here, we have taken  the order one parameters  to be  $a-e=0.7\sim 1.3$ and the FN parameter
$\lambda=0.18\sim 0.22$. We have found that
the predicted  $\sin^2\theta_{13}$ and  $\sin^2\theta_{12}$ are also completely consistent with
the experimental data.
Our numerical results depend on the scanning region  $a-e=0.7\sim 1.3$.
The condition of  ${\rm  det} [M_\nu]=0$  is  essential for the nontrivial prediction of $\delta_{CP}$. 
The allowed region of $\delta_{CP}$  is consistent with the recent T2K and NO$\nu$A data.
The CP conservation  $\delta_{CP}=0,\pm\pi$ is excluded.

In order to see the effect of the order one parameters $a-e$ on  our prediction of $\delta_{CP}$,
we present the frequency distributions of $\delta_{CP}$ for   $a-e=0.5\sim 2$ in Fig.10. 
 As the region of the parameter $a-e$ expands, the frequency distribution becomes broader.
  Notice that the hierarchies in the neutrino mass matrix $M_\nu$ predicted by the FN mechanism
 becomes obscure with such a large region of the parameters $a-e$ as stressed in section 3.
 In conclusion, we claim that ${\rm  det} [M_\nu]=0$  predicts  $\delta_{CP}$ as seen in Fig.6
 if the FN  flavor structure is sharp.
 
 It is helpful to comment on why ${\rm  det} [M_\nu]=0$ rules out $\delta_{CP}=0,\pm \pi$ as seen in Fig. 6.
 The five neutrino  experimental data,  two mass squared differences and three mixing angles,
 are possibly reproduced by six parameters $a-e$ and $m_0$ 
 without complex phases because of enough number of free parameters. Then, neutrino sector is the CP conserved one.
 When ${\rm  det} [M_\nu]=0$ is imposed, we have  five real parameters, 
 and so we cannot reproduce the experimental data if  $a-e$ are constrained around $1$ without the  CP violating phase.
 Thus, the CP conserved case $\delta_{CP}=0,\pm \pi$  is ruled out by the condition ${\rm  det} [M_\nu]=0$.

The condition of ${\rm  det} [M_\nu]=0$ is derived easily 
by assuming  two families of  heavy  right-handed neutrinos in the framework of the seesaw mechanism.
Notice that the neutrino mass matrix $M_\nu$ in Eq.(\ref{Mnu}) is determined only by the FN charges of the left-handed leptons after the integration of the right-handed neutrinos.

It is emphasized that the scenario with the two family heavy right-handed neutrinos is not necessarily required.
In practice,  we have checked 
that our prediction of $\delta_{CP}$ is not changed  in the case of  $m_1$ being smaller than $10^{-4}$eV.
However, we do not address the model with tiny $m_1$ since it is beyond the  scope of our work.

We have also found the remarkable  correlation  between  $\delta_{CP}$ and $\sin^2\theta_{23}$.
If  $\sin^2\theta_{23}$  is  larger than $0.5$,  $\delta_{CP}$ converges to around $\pm \pi/2$.
We expect the accurate measurement of  $\sin^2\theta_{23}$ will be done in near future experiments.
The  effective mass  in the  neutrinoless double 
beta decay $m_{ee}$ is also predicted to be $m_{ee}=3.3-4.0$ meV.

We should note that our results are consistent with the conclusions in \cite{Rink:2016knw},
where an exchange symmetry between two heavy right-handed neutrinos is  further imposed.
The CP violating phase $\delta_{CP}$ is predicted near by the maximal value $\pm \frac{\pi}{2}$ due to the
exchange symmetry.

\vspace{1 cm}
\noindent
{\bf Acknowledgement}

  We thank G. Branco for careful reading of the manuscript.
This work is supported by JSPS Grants-in-Aid for Scientific Research
15K05045, 16H00862 (MT) and 26287039, 26104009, 16H02176 (TTY).
 This work receives a support at IPMU by World Premier International Research Center
Initiative of the Ministry of Education in Japan.

\end{document}